\documentclass[sigconf]{acmart}
\usepackage{glossaries}
\usepackage{acronym}
\usepackage{hyperref}
\usepackage[noabbrev,capitalize]{cleveref}
\usepackage{csquotes}

\AtBeginDocument{%
  \providecommand\BibTeX{{%
    \normalfont B\kern-0.5em{\scshape i\kern-0.25em b}\kern-0.8em\TeX}}}

\copyrightyear{2022} 
\acmYear{2022} 
\setcopyright{acmlicensed}\acmConference[MODELS '22 Companion]{ACM/IEEE
25th International Conference on Model Driven Engineering Languages and
Systems}{October 23--28, 2022}{Montreal, QC, Canada}
\acmBooktitle{ACM/IEEE 25th International Conference on Model Driven
Engineering Languages and Systems (MODELS '22 Companion), October
23--28, 2022, Montreal, QC, Canada}
\acmPrice{15.00}
\acmDOI{10.1145/3550356.3561576}
\acmISBN{978-1-4503-9467-3/22/10}

\begin{document}

\title[MDE4ML: MontiAnna vs. ML-Quadrat]{MDE for Machine Learning-Enabled Software Systems: A Case Study and Comparison of MontiAnna \& ML-Quadrat}

\author{Jörg Christian Kirchhof, Evgeny Kusmenko, Jonas Ritz, Bernhard Rumpe}
\affiliation{%
  \department{Software Engineering}
  \institution{RWTH Aachen University}
  %\city{Aachen}
  \country{Germany}
  }
  \email{www.se-rwth.de}

\author{Armin Moin}
\affiliation{%
  \department{School of Computation, Information and Technology}
  \institution{Technical University of Munich, University of Antwerp \& Flanders Make}
  \country{Germany \& Belgium}}
  \email{armin.moin@tum.de}
  
\author{Atta Badii}
\affiliation{%
  \department{Department of Computer Science}
  \institution{University of Reading}
  \country{United Kingdom}}
  \email{atta.badii@reading.ac.uk}

\author{Stephan Günnemann}
\affiliation{%
  \department{School of Computation, Information and Technology}
  \institution{Technical University of Munich \& \\
  Munich Data Science Institute}
  \country{Germany}
  }
  \email{guennemann@in.tum.de}

\author{Moharram Challenger}
\affiliation{%
  \department{Department of Computer Science}
  \institution{University of Antwerp \\
  \& Flanders Make}
  \country{Belgium}
  }
  \email{moharram.challenger@uantwerpen.be}
\thanks{This research has partly received funding from the Federal Ministry for Economic Affairs and Climate Action (BMWK) in a project called KI-LaSt under grant no. 19I21036F. The responsibility for the content of this publication is with the authors.}
\renewcommand{\shortauthors}{Kirchhof et al.}


% Introduction: Main Topic (2 sätze)
% Key research question (1 Satz)
% How do you tackle the research question/ What’s your big new idea (1 Satz)
% Why nobody else has answered this question (2 Sätze)
% How did you go about doing the research? What’s the key impact of your research? (1 Satz)

\begin{abstract}
In this paper, we propose to adopt the MDE paradigm for the development of \gls{ML}-enabled software systems with a focus on the \gls{IoT} domain. We illustrate how two state-of-the-art open-source modeling tools, namely MontiAnna and ML-Quadrat can be used for this purpose as demonstrated through a case study. The case study illustrates using \gls{ML}, in particular deep Artificial Neural Networks (ANNs), for automated image recognition of handwritten digits using the MNIST reference dataset, and integrating the machine learning components into an \gls{IoT}-system. Subsequently, we conduct a functional comparison of the two frameworks, setting out an analysis base to include a broad range of design considerations, such as the problem domain, methods for the \gls{ML} integration into larger systems, and supported \gls{ML} methods, as well as topics of recent intense interest to the \gls{ML} community, such as AutoML and MLOps. Accordingly, this paper is focused on elucidating the potential of the MDE approach in the \gls{ML} domain. This supports the \gls{ML}-engineer in developing the (ML/software) model rather than implementing the code, and additionally enforces reusability and modularity of the design through enabling the out-of-the-box integration of \gls{ML} functionality as a component of the \gls{IoT} or cyber-physical systems.
\end{abstract}




\maketitle
%\input{Glossaries}
% 20. Juli Deadline

\section{Introduction}\label{sec:intro}
Model-Driven Engineering (MDE) aims to use models to support the development of engineered systems (such as software) at various stages, for example, through generating implementations from the models \cite{mde}. Further, \gls{ML} is a branch of Artificial Intelligence (AI), currently of high impact, mostly due to the rise of deep learning technologies, to enable machines to \textit{learn} through inference on data \cite{lecun2015deep}. \gls{ML} is used today in almost all areas and domains of software engineering. It is a popular choice for addressing problems that are difficult to overcome using traditional programming. In traditional programming, the developer has to specify the solution to the problem in an imperative or declarative manner. In contrast, \gls{ML} is useful when the problem is highly data-intensive and the pattern space is at a large scale and too complex for human beings to analyze and solve directly. In such cases, inferences from observed data can be useful. However, \gls{ML} is often not the best approach to follow in situations where the solution can be realized directly (i.e., in the traditional programming way) rather than through inference \cite{Leskovec+2014}. In this work, we focus on smart software systems that are not rule-based (e.g., expert systems), but require inference from observed data, such as data collected though wireless sensors, also referred to as ambient intelligence, particularly in the context of smart environments, such as smart home, smart city, and smart mobility applications.

It transpires that creating the \gls{ML} components for such software systems is often highly challenging for software developers. Even when using the high-level APIs of \gls{ML} frameworks and libraries, such as Scikit-Learn \cite{Pedregosa+2011}, TensorFlow \cite{Abadi+2015}, Keras \cite{Chollet+2015}, or MXNet \cite{Chen+2015}, developers need to have knowledge both about \gls{ML} itself, and about the framework used in order to achieve satisfactory results. However, many tasks in the creation of \gls{ML} pipelines, such as the addition of a convolutional layer to a deep neural network model, often follow a common set of steps, and differ essentially in the (hyper)parameters tuning phase.

Model-driven development promises to solve the above-mentioned problems in an efficient manner by raising the level of abstraction, and providing automation, for example, for code generation based on high-level specifications. Therefore, in addition to the above-mentioned \gls{ML} libraries, a number of model-driven tools and associated domain-specific languages, such as MontiAnna \cite{KNP+19,AKK+21} and ML-Quadrat \cite{Moin+2022-SoSyM, ML-Quadrat} have been introduced in recent years. These frameworks utilize high-level specifications of \gls{ML} models (e.g., deep learning networks) to generate code according to an underlying \gls{ML} framework. By raising the level of abstraction, they liberate the developers from having to learn the specifics of the \gls{ML} framework, and enable them to focus on the selection of the \gls{ML} solutions they desire instead.

One application area where \gls{ML} has proven useful is the \gls{IoT} domain, which is suitable for the application of \gls{ML} at several levels: On the one hand, \gls{IoT} applications must be able to make inferences about their environment based on sensor data\footnote{When we use the term \emph{sensor data}, we also refer to data from highly complex sensors, such as cameras.} (i.e., ambient situation assessment). Due to the data (stream) volume, velocity, and the complexity of the data, it can be difficult to derive knowledge about the environment from the data. This is where \gls{ML} can help to identify patterns in the data, for example, to recognize a particular face in a photo, and consequently infer that a particular person is at a location. On the other hand, \gls{ML} can also be used to make predictions about the future based on existing data. For example, patterns in the behavior of smart home residents can be detected to predict patterns of occupancy in rooms or spaces, and accordingly schedule cleaning tasks by a domestic robot cleaner.

In this paper, we study and compare two state-of-the-art model-driven \gls{ML} tools, namely MontiAnna and ML-Quadrat, particularly from an \gls{IoT} perspective, with a focus on their functional aspects. Accordingly, this paper makes the following contributions: i) It presents a case study to illustrate using both tools for modeling \gls{ML}-enabled software. It also briefly studies each tool from an \gls{IoT} perspective. ii) It conducts a functional comparison between the two model-driven tools. iii) It illustrates how \gls{mde} can support different aspects of the development of machine learning-driven software. 

\begin{figure}
	\centering
	\includegraphics[width=0.7\columnwidth]{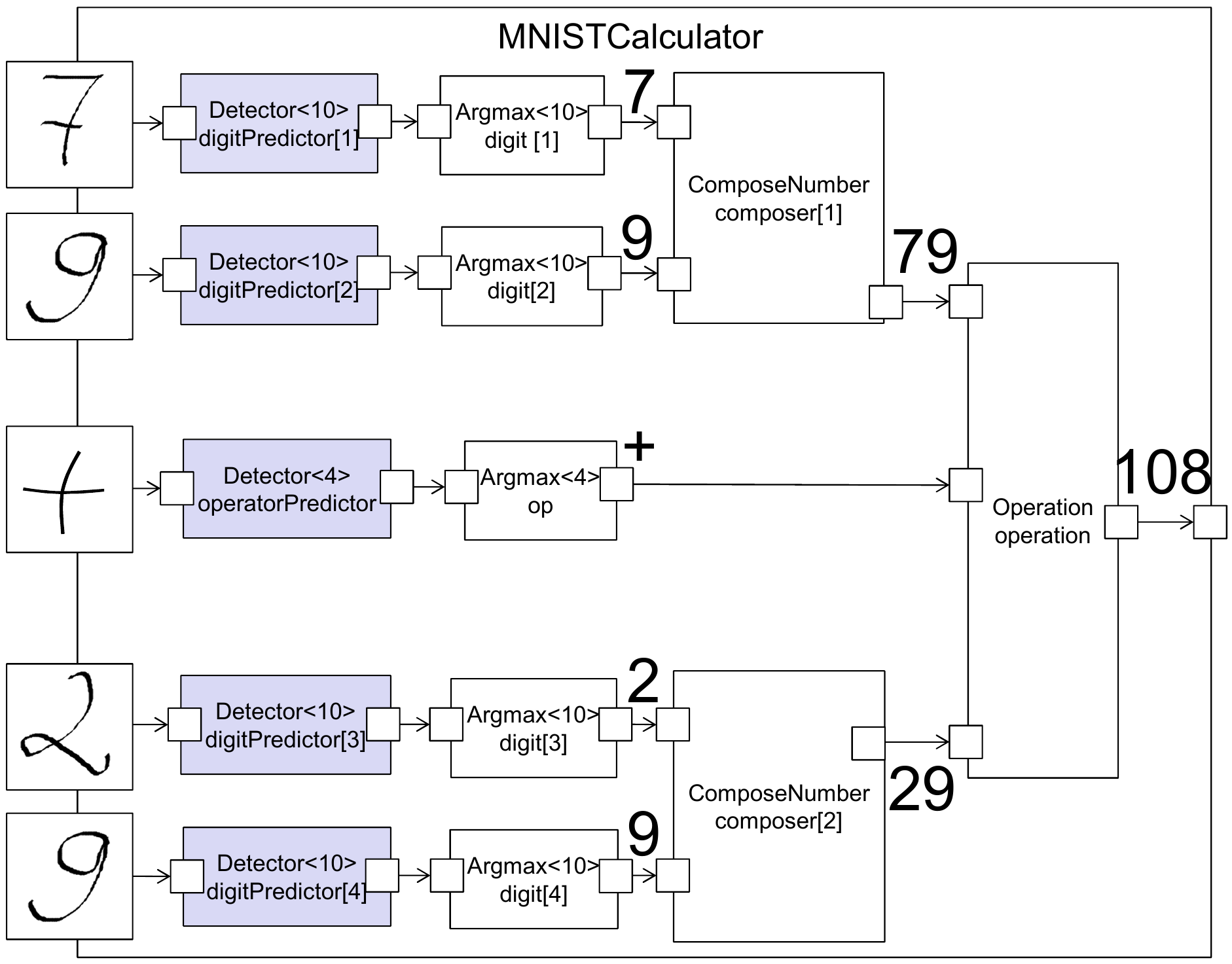}
	\caption{MNIST Calculator taking hand-written digits as inputs and outputting the sum according to the dataflow \cite{KNP+19,Kus21}}
	\label{fig:mnistcalc}
\end{figure}

As a common ground, we use the MNIST Calculator model from \cite{KNP+19,Kus21}, which is a \lq{}hello world\rq{}-like example for the so-called \textit{software 2.0} applications, consisting of \gls{ML} and standard software components. The architecture of the MNIST Calculator is given in \cref{fig:mnistcalc}.

The remainder of this paper is structured as follows. Section \ref{sec:preliminaries} introduces the MontiAnna and ML-Quadrat tools through a case study. Moreover, an in-depth comparison between MontiAnna and ML-Quadrat, which is mostly concentrated on their functional aspects is provided in Section \ref{sec:comparison}. Further, we review the related work in Section \ref{sec:relatedwork}. Finally, we conclude in Section \ref{sec:conclusion-future-work}. %1
\section{Tools presentation and case study}
\label{sec:preliminaries}
In this section, we present both modeling tools and demonstrate how to use them through a case study for image recognition of handwritten digits. Hence, this section conforms to the first contribution mentioned in Section \ref{sec:intro}. 

\subsection{MontiAnna}
\label{subsec:MontiAnna}
MontiAnna \cite{KNP+19} is a textual modeling framework and build system for the design of deep \glspl{ann}. It is embedded into the component and connector-oriented modeling language family EmbeddedMontiArc \cite{KRRW17,KRSvW18a,KKR19}, where the software functionality is encapsulated into components with interfaces defined by input and output ports. The communication of the components needs to be modeled explicitly by creating connectors between ports. A MontiAnna \gls{ann} is integrated into larger software architectures using the same component-based principle, which means each neural network is a component \cite{KPRS19}. The component's ports are then mapped to the input and output layers of the network, respectively. The MontiAnna build system detects all \glspl{ann} in a software architecture and resolves trained model weights for these networks at build time. If no trained weights are available for a given network, or if the network model or the designated training data have changed, the build system trains the network automatically. The developer does not have to deal with the \gls{ML} life cycle manually. An \gls{ML} model in MontiAnna consists of a neural network architecture encapsulated into a component hull and a separate configuration model setting up the desired training scheme and holding the hyperparameters. Furthermore, tag models can be used to assign versionable training data and/or existing weights to the \glspl{ann}. 

\begin{figure}
	\centering
	\includegraphics[width=0.9\columnwidth]{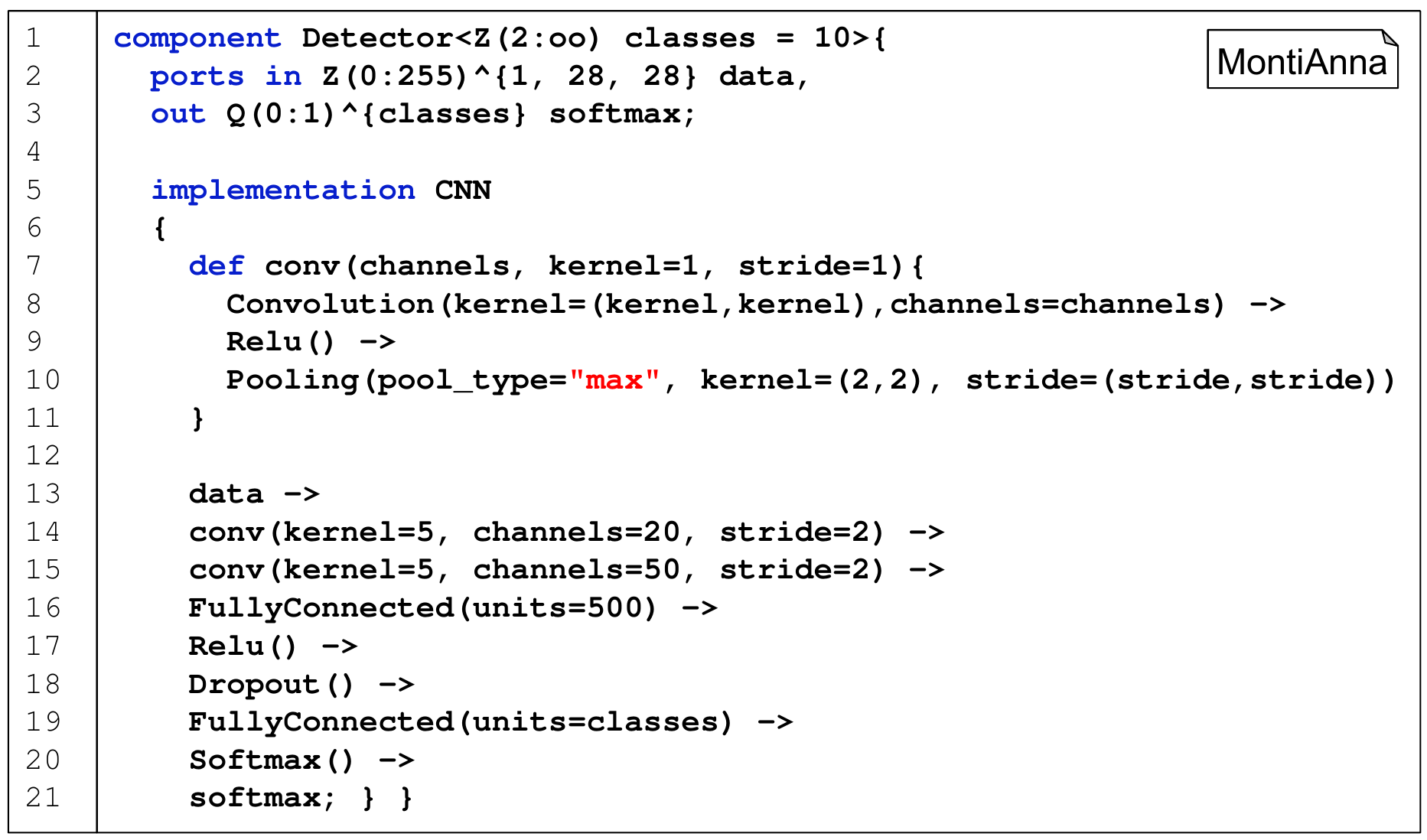}
	\caption{MNIST Detector CNN modeled in MontiAnna.}
	\label{fig:mnist_detector}
\end{figure}

MontiAnna provides a series of out-of-the-box training pipelines, for example, for supervised learning, several variants of reinforcement learning \cite{GKR19}, \glspl{gan}, and \glspl{vae}. Further pipelines or pipeline components can be implemented as standard Python code and assembled using the component-based paradigm. To keep track of the available configuration options for the different pipelines, MontiAnna uses a modular schema system, enabling inheritance and combining of the configuration parameters. \Cref{fig:mnist_detector} shows a MontiAnna deep learning component encapsulating a \gls{cnn} for the detection of MNIST digits.

%%MNIST Detector Beispiel
%%https://git.rwth-aachen.de/monticore/EmbeddedMontiArc/applications/mnistcalculator/-/blob/master/emadl-maven-plugin/src/main/emadl/calculator/Network.emadl

In the context of the \gls{IoT}, MontiAnna can be used as part of MontiThings, which is a DSL-based framework \cite{KRS+22} to specify the behavior of components in \gls{IoT} applications. For example, even if edge devices in smart home applications may indeed have insufficient computational resources to perform full speech recognition themselves, many of them could detect a key phrase (e.g., the wake up commands \enquote{Hey Siri} or \enquote{OK Google}) to know when to start recording. Using MontiAnna, models for such simplified speech recognition can be trained and integrated into \gls{IoT} applications.

\Cref{fig:mnist_detector} shows a MontiAnna component definition specifying the \gls{ann} architecture for the MNIST detector components of the MNIST calculator architecture of \cref{fig:mnistcalc}. The component is named \texttt{Detector} in line 1 in \cref{fig:mnist_detector}. Furthermore, it is defined as a generic component with the generic interface parameter \texttt{classes}. This supports the usability of the network architecture for similar problems with a different number of classes (with appropriate re-training). The interface is defined in lines 1-2 in \cref{fig:mnist_detector}, streaming the input data for 28x28 matrices within the range of 0 and 255 for grayscale images. The output is a softmax vector with the dimensionality defined by the \texttt{classes} parameter. 

The layer-specific \gls{ann} architecture is defined in the implementation block (see lines 13-21). The input and output ports' names are used as the input and output of the network (see lines 13 and 21, respectively). The network layers are instantiated by using library layers such as \texttt{Convolution}, \texttt{FullyConnected}, etc. Reoccurring patterns are grouped to new layer classes in the \texttt{def} block in lines 6-11. Should the user be uncertain as to the how to select the appropriate architecture, wild card layers could be included; whereby the framework includes network layers iteratively based on a heuristic, e.g. AdaNet \cite{cortes2017adanet}, compares the results of each iteration, and eventually returns the best \gls{ann} model found. The same network architecture can be re-used to learn an operator detector for the MNIST calculator by providing the needed training examples and changing the class parameter to the number of supported operators, for example, +, -, $\times, \div$. 
The machine learning components can then be interconnected with other components using connectors as was shown in \cite{KNP+19,KRSvW18a}.

\subsection{ML-Quadrat}
ML-Quadrat is a modeling tool for creating \gls{ML}-enabled \gls{IoT} services. It offers full code generation out of software models and comes with a desktop version, as well as a web-based version. It is based on the Eclipse Modeling Framework (EMF) \cite{EMF} and the Xtext framework \cite{Xtext}. It has a text-based model editor, which uses Xtext and offers typical IDE features, such as syntax highlighting and auto-completion, as well as a tree or form-based model editor, which is based on the EMF tree model editor. ML-Quadrat generates Python and Java code. The Python code is responsible for the \gls{ML} components and uses the APIs of Scikit-Learn or Keras with the TensorFlow backend. The generated Python and Java codes are seamlessly integrated. In addition, ML-Quadrat inherited the DSL of the modeling tool ThingML \cite{Harrand+2016, ThingML}, including its code generation framework. Thus, it generates code for a range of programming languages and platforms, for example, C code for the Arduino and POSIX platforms. 

A software model in ML-Quadrat consists of a structural part, a behavioral part, an ML component, as well as a configuration part. The latter is absent in the case of Platform-Independent Models (PIMs). However, for code generation, Platform-Specific Models (PSMs), which include configurations, are needed \cite{Moin+2022-COMPSAC}. Note that \textit{configurations} in ML-Quadrat are concerned with the entire software model, thus these are different from the \textit{configurations} in MontiAnna. The structural part specifies the \textit{things} in the system, as well as their \textit{ports}, \textit{messages}, \textit{parameters}, \textit{properties} (i.e., local variables), etc. A \textit{Thing} in ML-Quadrat can be seen as an \textit{actor} or \textit{agent} that is connected to the \gls{IoT}. \textit{Things} can communicate with each other asynchronously through message-passing via their \textit{ports}. Furthermore, each \textit{thing} must have a behavioral part, which is a Finite-State Machine (FSM), also known as a state machine, state diagram, or state chart. The semantics are adopted from the UML 2 standard \cite{OMG-UML-2017}. It is this part of the software model that deploys the ML model predictions to make the software model \textit{smart}. For instance, the predictions of the ML model may affect the state transitions. 

\begin{figure}
	\centering
	\includegraphics[width=0.9\columnwidth]{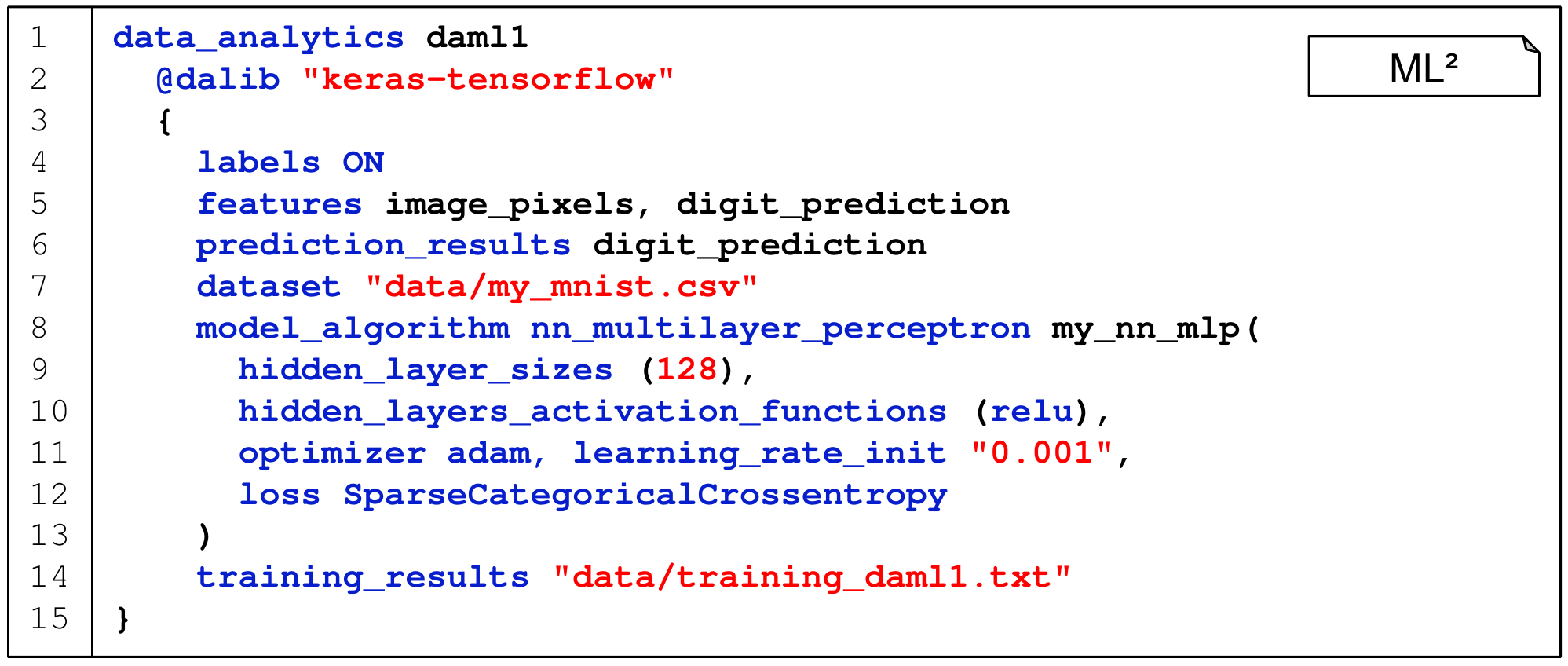}
	\caption{MNIST detector MLP \gls{ann} modeled in ML-Quadrat.}
	\label{fig:ml-quadrat-mnist-model-daml-part}
\end{figure}

The \gls{ML} part of the software model typically comprises the specific \gls{ML} method that should be deployed, for instance, the model architecture family (e.g., the Multi-Layer Perceptron \gls{ann}), the hyperparameters, such as the choice of the optimizer or learning algorithm for training the \gls{ML} model, the number of hidden layers, the layer sizes, and the choice of the activation functions in each layer in the case of \glspl{ann}. If a hyperparameter is absent in the software model, its default value in the respective \gls{ML} library or framework (e.g., Keras/TensorFlow) will be assumed. In the case that an \gls{ML} method is available in both Scikit-Learn and Keras, the practitioner may explicitly select the one to be used, or the system can decide on its own. Moreover, the practitioner may bring a pre-trained \gls{ML} model, which has any arbitrary architecture and has been trained with any arbitrary algorithm and connect it to the software model. This brings some flexibility since the options for \gls{ML} methods will not be limited to the ones that are supported out of the box. Figure \ref{fig:ml-quadrat-mnist-model-daml-part} illustrates the \gls{ML} part of the textual model instance in ML-Quadrat that models a software service for automated handwritten digit recognition based on the MNIST reference dataset using an MLP. The DSL keywords are highlighted in blue. The annotation \textit{da\_lib} enables the practitioner to select the target \gls{ML} library. For instance, we support the APIs of both Scikit-Learn and Keras (with the TensorFlow backend) for the MLP \gls{ann} method. Moreover, the \textit{labels} keyword specifies whether the data are class labeled (i.e., supervised learning is applicable) or not. The possible options here are \textit{ON}, \textit{OFF}, and \textit{SEMI} for supervised, unsupervised, and semi-supervised learning, respectively. Moreover, the \gls{ML} features are listed after the \textit{features} keyword. Further, the \textit{prediction\_results} keyword can specify where the future predictions of the \gls{ML} model should be stored once the training is accomplished. Additionally, the \textit{dataset} keyword is used to introduce the path of the Comma-Separated Values (CSV) file that contains the training data on the file system. The core of this part of the software model is the \textit{model\_algorithm} specification, which models the \gls{ML} method that should be created and deployed. In this case study, we train an MLP \gls{ann}. The hyperparameters are mostly selected from the respective \gls{ML} libraries, such as Scikit-Learn. If a hyperparameter (e.g., optimizer) is missing, the default choice of the respective library will be selected automatically. In this example, only one hidden layer with the size of $128$ will be created. For instance, if another hidden layer of size $64$ had been desired, we would have (128, 64) instead of (128) for the \textit{hidden\_layer\_sizes} hyperparameter. Likewise, the activation function of each hidden layer needs to be specified through the \textit{hidden\_layers\_activation\_functions} hyperparameter respectively. Furthermore, the rest of the shown hyperparameters specify the optimizer (in this case Adam), the initial learning rate, as well as the choice of the loss function. Finally, the training log will be stored in the stated text file whose path is provided through the \textit{training\_results} keyword. 

\begin{figure}
	\centering
	\includegraphics[width=0.7\columnwidth]{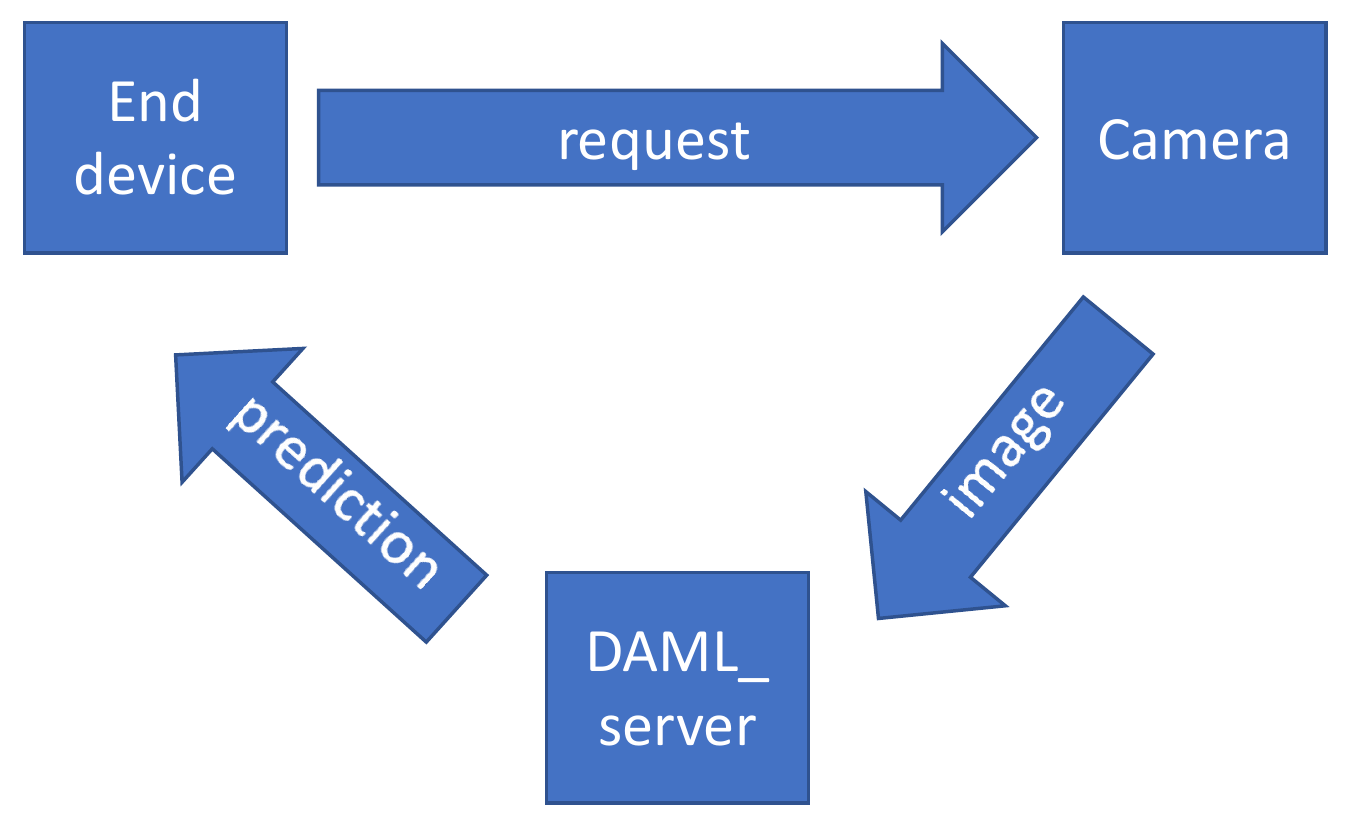}
	\caption{Architecture of the generated service via ML-Quadrat for handwritten digit recognition.}
	\label{fig:ML_Quadrat_service_arch}
\end{figure}

Figure \ref{fig:ML_Quadrat_service_arch} depicts the overall architecture of the \gls{IoT} service that deploys the mentioned \gls{ML} components in order to offer automated handwritten digit recognition. This service comprises three \textit{things}, which are represented by the blue rectangles: i) An end-device (such as a smartphone or tablet); ii) A camera; iii) A server for carrying out the Data Analytics and \gls{ML} (DAML) tasks, called DAML\_server. The \gls{ML} component shown in Figure \ref{fig:ml-quadrat-mnist-model-daml-part} belongs to the latter.

\begin{figure}
	\centering
	\includegraphics[width=0.9\columnwidth]{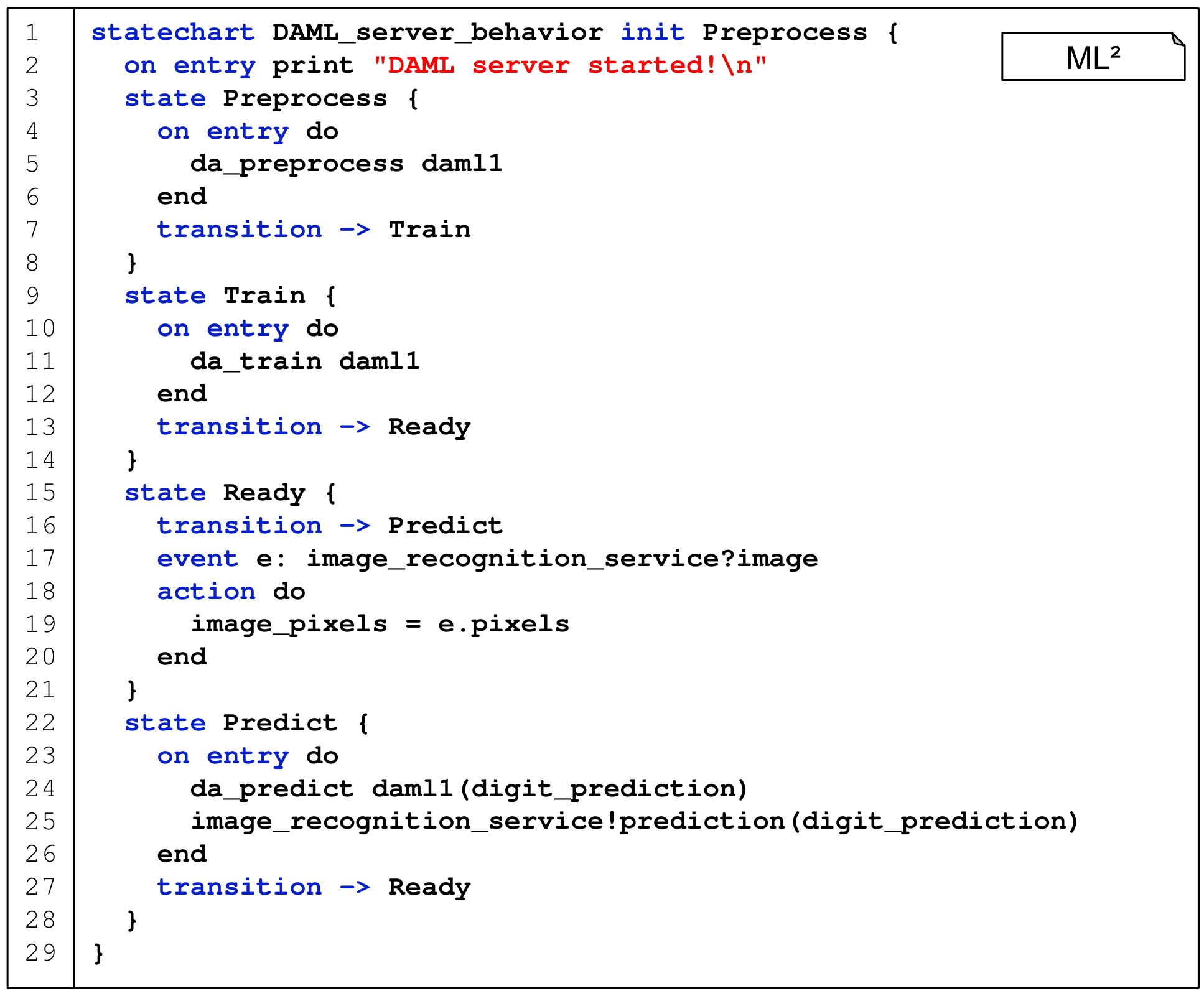}
	\caption{The behavioral model of DAML\_server in ML-Quadrat.}
	\label{fig:ml-quadrat-mnist-model-fsm-part}
\end{figure}

Finally, the behavioral model of the DAML\_server \textit{thing}, which deploys the above-mentioned \gls{ML} component, as shown in Figure \ref{fig:ml-quadrat-mnist-model-daml-part}, is presented in Figure \ref{fig:ml-quadrat-mnist-model-fsm-part}. Initially, the data pre-processing is conducted. Then, the \gls{ML} model is trained. Next, the system will switch to the ready state, which is the standby state. Once an image is received on the \textit{image\_recognition\_service} port, its pixel intensities will be provided to the \gls{ML} model to recognize the digit. Afterwards, the system reverts to the ready state, thus standing by again. Note that the \textit{da\_preprocess}, \textit{da\_train}, and \textit{da\_predict} \textit{actions} lead to the execution of the data pre-processing (i.e., data preparation), \gls{ML} model training, and prediction tasks of the \gls{ML} pipeline. Lastly, the question mark and the exclamation mark, which are used in the statechart in Figure \ref{fig:ml-quadrat-mnist-model-fsm-part} check for receiving a particular message on a specific port, and result in sending a particular message on a specific port, respectively.

 %2 
\section{Comparison}
\label{sec:comparison}
Table \ref{tab:comparison} summarizes a comparison of MontiAnna and ML-Quadrat with respect to several aspects of their capabilities. Below, we elaborate on these by reference to the respective rows. This section constitutes the second and the third contributions mentioned in Section \ref{sec:intro}, as on the one hand it shows a functional comparison of the two model-driven tools, while on the other hand it elucidates the broad range of possible support through \gls{mde} in the development of \gls{ML}-driven software.

\textbf{Problem domain and Integration:} The first aspect to be compared in examining the two modeling tools is the specific problem domain on which they focus. MontiAnna offers the possibility to create standalone \gls{ML} applications, while the \gls{ML} functionality can also be used in the \gls{IoT} domain by using MontiThings \cite{KRS+22}. Therefore, the \gls{ML} functionality may be encapsulated into a component and re-used in the EmbeddedMontiArc framework for integration into the full system, or re-used in MontiThings for integration into \gls{IoT} services. By contrast, ML-Quadrat offers an integrated modeling language, which covers both IoT and ML. In fact, this provides for a one-stop-shop for \gls{ML}-enabled heterogeneous and distributed services for the \gls{IoT}. This is realized by the integration of the APIs of the \gls{ML} libraries and frameworks Scikit-Learn and Keras into the prior work ThingML \cite{Harrand+2016, ThingML} both at the meta-model level and at the level of code generators.

\textbf{Modeling methodology and \gls{ML} methods:} In addition, the tools differ concerning the modeling methodology and the corresponding \gls{ML} methods. MontiAnna is concerned with a particular family of \gls{ML} models, namely deep \glspl{ann}. Accordingly, MontiAnna has its dedicated modeling language for \glspl{ann} , called \texttt{CNNArchLang}, which provides for layer-specific modeling of \glspl{ann} with a broad range of supported layers, as shown in section \ref{subsec:MontiAnna}. The network architecture is automatically validated, such that - among other checks - it is ascertained that the output dimensions and input dimensions of consecutive layers match. In MontiAnna, the networks can be trained through a supervised learning approach, a reinforcement learning attempt, and unsupervised learning approaches, such as \glspl{gan} or \glspl{vae}. Moreover, there is a strict differentiation between the network architecture and the hyperparameters. A dedicated language called \texttt{ConfLang} was developed, which has a JSON-like syntax for configuring a broad range of hyperparameters. These include the number of epochs to be trained, the batch size, and even a nested configuration establishing the optimizer and additionally determining its parameters such as the learning rate. Distinguishing between the network architecture modeling language and a language for hyperparameters builds a clear separation of concerns and avoids mixing up domains. In contrast, ML-Quadrat offers a single DSL for modeling the entire software service or application, including the ML model architecture, the hyperparameters for ML model training, as well as other elements (e.g., \gls{IoT} components). ML-Quadrat supports two ways of deploying machine learning methods. The practitioner may either select an ML method that is supported out of the box (e.g., MLP \gls{ann}, or decision trees), or use a pre-trained ML model. In the latter case, which is called the \textit{blackbox} ML mode (given that the ML model is dealt with as a black-box by the software model), the ML model may possess an arbitrary architecture in the supported libraries and could be trained using any arbitrary learning algorithm, method, and techniques, which are supported in the libraries. Hence, the practitioner's options are not limited to the out-of-the-box ML methods. These include linear methods for classification and regression (i.e., logistic regression and linear regression), Na\"ive Bayes with various kernels (e.g., Gaussian and Bernoulli), Decision Trees, Random Forests, and MLP \glspl{ann} for supervised ML. In addition, K-Means, Mini-Batch K-Means, DB-SCAN, Spectral Clustering, and Gaussian Mixture Model are enabled for unsupervised ML. Last but not least, in the case of semi-supervised learning (i.e., partially labeled data), the Self-Training, Label Propagation, and Label Spreading methods can be deployed.

\textbf{Target \gls{ML} libraries \& frameworks and \gls{ML} pipelines:} When working with MontiAnna, the \gls{ML} engineer can configure a
\textit{backend}, which is an \gls{ML} library with a Python interface, to generate the code resulting from the model. So far, Caffe2, 
TensorFlow, Pytorch, and MxNet/Gluon are supported. This flexibility enables easy benchmarking between different backends and paves the way for 
experimentation with functionalities only supported in certain backends. The generated code concerning the network architecture and the training procedure is written in Python, while the execution can be done in both Python and C++, as C++ is the most common language in the target domain. However, ML-Quadrat supports the Python ML libraries Scikit-Learn, and Keras (with the TensorFlow backend). Besides the ML component, the rest of the \gls{IoT} services modeled in ML-Quadrat, may be generated for a range of target \gls{IoT} platforms, programming languages, and APIs. The choices include, but are not limited to C code for POSIX and Arduino, as well as Java, Javascript, and Go. If Java is desired, the generated Java code will be seamlessly integrated with the generated Python code for the ML component. To this aim, the Java and Python code generator that is offered by ML-Quadrat can be used.

\textbf{\gls{ML} pipelines:} Typically, \gls{ML} problems are tackled using an \gls{ML} pipeline (i.e., workflow). This pipeline can vary depending on the problem. However, it often incorporates some kind of pre-processing of the data (i.e., data preparation), a feature engineering step (usually for non-deep-learning approaches), an ML model training process (i.e., \textit{learning} the optimized parameters in the case of parametric ML models), and an evaluation, which ensures that no \textit{over-fitting} occurs. The pre-processing phase can consist of a change in the color space for images, data cleaning, imputation of missing values, normalization, standardization, and stratified sampling. In the case of End-to-End ML, the ML model itself includes all the pipeline implicitly and does everything self-sufficiently. Thus, when the input data are fed into the trained ML model, it knows all the necessary steps to deliver the final result (e.g., prediction). Modeling a pipeline implies supporting the practitioner at two levels: i) putting together components to create a pipeline; ii) creating the realization of the pipeline components. In MontiAnna, pipelines can be constructed using an established Component and Connector Language, called EmbeddedMontiArc. Components can be created and connected via ports. The realization of the components can take place via the following alternative routes: i) through selection out-of-the-box, for example, a data cleaning procedure provided by the framework; ii) the generation of the neural network and training procedure using the aforementioned capabilities of MontiAnna; iii) be handcrafted by the practitioner to suit the generated interfaces as derived from the components and their ports. In contrast, ML-Quadrat enables data pre-processing through the \textit{da\_preprocess} action of the DSL (see Figure \ref{fig:ml-quadrat-mnist-model-fsm-part} in Section \ref{sec:preliminaries}). Currently, this mostly comprises of feature scaling (i.e., standardization and normalization of numeric values), as well as label encoding (i.e., one-hot encoding) for categorical labels. The practitioner may either manually adapt the generated Python script for pre-processing or adapt the model-to-code transformation accordingly.

\textbf{Modularity and Compatibility:} Besides the modularity offered through the exchangeability of the pipeline components in MontiAnna, pre-trained networks can be loaded and imported as network layers. They are then trained together with the network architecture modeled around this layer. To guarantee framework interoperability, both export and import are implemented in the ONNX standard format. Furthermore, modularity in the context of ML-Quadrat is provided through the possibility of importing pre-trained ML models of various architectures from the supported ML libraries and (de)serialization of these models, for example, as Python Pickle objects for Scikit-Learn or HDF5 in the case of Keras.

\textbf{AutoML} AutoML is a rising \gls{ML} subtopic. In MontiAnna, the integration of AdaNet serves as the first step toward AutoML by aiming at conducting a Neural Architecture Search (NAS) for any problem. This approach seeks to create an \gls{ann} through an additive growing ensemble by putting together sub-networks. Moreover, with the help of the explained concept of pipelining, AutoML can easily be enabled. Therefore, MontiAnna possesses two levels of AutoML support: i) Automatically optimizing the parameters that define a component, for example, the hyperparameters of a training configuration. ii) At a higher level, complete components can be exchanged, if the exchanged component is taken from a set with suitable interfaces. Additionally, ML-Quadrat supports AutoML at the following two levels. First, it offers rule-based support by checking certain constraints based on the API documentation of the respective backend libraries, and ML domain knowledge. For instance, if a hyperparameter has been set outside the permitted or recommended range, the practitioner can be warned about this. Also, in certain cases, such as scaling numeric data in the data pre-processing of \glspl{ann}, or avoiding data shuffling and cross-validation in the case of sequential (e.g., time series) data, decisions will be made and enforced automatically should the AutoML mode be enabled. Second, for certain ML methods, automated ML model architecture/type selection, as well as automated hyperparameter optimization using Bayesian Optimization through the Hyperopt library can be offered. For the latter, the practitioner needs to use the standalone open-source tool AutoNIALM \cite{autonialm}, which was designed for a particular use case, namely energy disaggregation, but can be adapted and deployed for other problems as well.

\textbf{Re-training:} In MontiAnna, the training procedure is executed only if the input data or the model have changed. Otherwise, the trained model remains the same as compared to the previous training run. Automated re-training is also implemented in the MontiAnna framework: When an extension of an existing dataset is deployed, an event is triggered that initiates the re-training process of the model. The new training process starts, where the last training process ended and takes over the learned parameter, such as the learning rate that was automatically adapted in the previous run. By contrast, in ML-Quadrat, the practitioner may deploy a timer in the software model to re-train the ML models periodically, or it can occur in an event-based manner following the adopted event-driven programming paradigm (e.g., upon the receipt of a particular message type on a specific port of the \textit{thing} (i.e., the agent), which contains the respective ML component).

\textbf{Generated Artifacts and Artifact Management:} In MontiAnna, the generated artifacts comprise the source code for the creation of the system, the source code for the training of the ANN model, and either the out-of-the-box functionality for the pipeline components or pre-generated interfaces for the user to realize the pipeline component manually. When bundled as a package, these artifacts constitute the source code archive. Other archives being created by the framework are the ANN model archive, which includes the weights of the trained model, and the dataset archive, which contains the dataset associated with a connection to the ANN model which it was trained with. These packages can be managed with Apache Maven. Maven goals exist for the deployment of the archives as well as for the installation of the archives to the local machine. Similarly, ML-Quadrat generates all the artifacts of the software solution automatically out of the software model, which is designed by the practitioner. These include the entire source code, ML models (ANNs or other ML model families), as well as the build and run scripts. The generated source code is seamlessly integrated with the generated ML models and can train, deploy, and re-train them automatically as required. In the case of Java (and Python), this includes code generation, where the Python scripts are in charge of ML, and the Java code is responsible for the rest of the IoT service functionality; here, Apache Maven is deployed (similar to MontiAnna) for artifact and life-cycle management of the generated software solution. In this case, an executable JAR that contains all the dependencies of the generated IoT service will be produced, and can be used conveniently by the operator or end-user to deploy and run the IoT service.

\begin{table*}
\centering
\caption{MontiAnna vs. ML-Quadrat}
\begin{tabular}{|p{0.25cm}|p{2.5cm}||p{6.5cm}|p{6cm}|  }
 \hline
 & \textbf{Properties} & \textbf{MontiAnna} & \textbf{ML-Quadrat} \\
 \hline
 1 & Problem domain & ML self-contained (or together with MontiThings for the IoT) & ML-enabled \gls{IoT} services \\
 \hline
 2 & Integration & Component encapsulation in EmbeddedMontiArc / MontiThings  & Encapsulation in ThingML \\
 \hline
 3 & Modeling methodology & Multiple DSLs & Single DSL \\
 \hline
 4 & ML methods & Various deep \glspl{ann} with supervised, unsupervised and reinforcement learning & Various supervised, unsupervised, and semi-supervised ML approaches \\
 \hline
 5 & Target ML libraries \& frameworks & MXNet Gluon, TensorFlow, PyTorch, Caffe2 & Scikit-Learn and Keras (with the TensorFlow backend) \\
 \hline
 6 & Target languages for code generation & Python (training and runtime), C++ (runtime) & Python (for ML), Java, C, Javascript, and Go \\
 \hline
 7 & ML pipelines & Modeled via the DSL; functionality out of the box or implemented by user & Out-of-the-box \\
 \hline
 8 & Modularity and Compatibility & Pre-trained networks can be loaded as network layers and ONNX \cite{ONNX} & Any pre-trained ML model in Scikit-Learn or Keras may be imported/plugged in\\
 \hline
 9 & AutoML & Neural Architecture Search with AdaNet, hyperparameter optimization and component-based AutoML planned & Rule-based and through Bayesian Optimization (Hyperopt). \\
 \hline
 10 & Re-training & Re-training only if data or model have changed, event-based automated re-training & Event-based automated re-training, for example, when new data arrive, or timer-based. \\
  \hline
 11 & Generated artifacts & Full source code containing \glspl{ann} (from \texttt{CNNArchLang} models) and ML model training scripts (from \texttt{ConfLang}), complete source code and dataflows or only interfaces (from Pipeline Model created with \texttt{EmbeddedMontiArc} and \texttt{ConfLang}) & ML models, full source code (including the Python scripts for pre-processing, training, and prediction), build and run scripts. \\
 \hline
 12 & Artifact management & Maven-based artifact re-use (source code, trained models, datasets are packaged independently) & Apache Maven projects generated for Java \\
 \hline
\end{tabular}
 \label{tab:comparison}
\end{table*}
%%mit beispiel $2$
%\input{tex/04.VerdictAndDiscussion.tex}%%futurework %1
\section{Related Work}
\label{sec:relatedwork}
Various prior works in the literature addressed the topic of deploying high-level specifications, abstractions, and visual programming, to improve AI (in particular \gls{ML}) engineering. In the following, we briefly review some of them them. We are particularly interested in those which deployed the \gls{mde} paradigm.

First, high-level APIs concerning \gls{ML} were provided through the \gls{ML} frameworks, such as Scikit-Learn \cite{Pedregosa+2011}, TensorFlow \cite{Abadi+2015}, Torch \cite{Torch} and Keras \cite{Chollet+2015}. More of these frameworks are presented in \cite{KNP+19}. These frameworks come with methods to build, train, and evaluate neural networks as well as other \gls{ML} models. Usually, they are implemented in C or C++ and accessed via a Python or C++ API. Although they are very comprehensive, the \gls{ML} engineer has to implement their solutions with a general-purpose language, thus being obliged to learn it for each and every platform that is needed beforehand. Second, \gls{ML} workflow designers and workbenches, such as KNIME \cite{Berthold+2009}, WEKA \cite{Hall+2009}, and RapidMiner \cite{RapidMiner} aimed for supporting a more efficient \gls{ML} practice. In addition, visual programming for \gls{ML} was enabled through a number of tools, such as TensorBoard \cite{TensorBoard}. Further, Infer.Net \cite{Bishop2013, InferNet} proposed \gls{mde} for \gls{ML}. However, they were focused on probabilistic programming, thus using Probabilistic Graphical Models (PGMs) as software models for producing the entire software source code in C\# out of them. Moreover, GreyCat \cite{Hartmann+2017,Hartmann+2018,Hartmann+2019} seamlessly integrated \gls{ML} into domain models. Their work was similar to ML-Quadrat, but only targeted Java, Javascript, and Typescript for code generation. Thus, it was not suitable for typical resource-constrained \gls{IoT} platforms. Another MDE solution for the deep learning domain used in practice is ML.NET, which was developed by Microsoft. The framework promises \textit{"authoring production-grade machine learning pipelines [...]"} \cite{ahmed2019machine}. Based on the .NET platform, developers can create pipelines as Directed Acyclic Graphs (DAG) with out-of-the-box functionalities that are easy to share efficiently through an abstraction called DataView. However, the model customizability is limited to the parameters. Furthermore, IBM SPSS Neural Networks \cite{al2019gateway} offers the possibility to integrate neural networks to IBM SPSS, which is common a software for statistics and data analytics. This approach can simultaneously be seen as a reduction of the manual expense to the minimum and a realization of AutoML. Instead of designing the neural network architecture manually, it is seen as a non-configurable black box simply automatically created based on the data. As IBM SPSS is statistical software, it does not support convolutional layers for images or graph convolutional layers, but the fully-connected architecture model aims at, for instance, solving classical statistical tasks using neural networks. These can thereby serve as a surrogate for linear regression. Finally, Azure Machine Learning is a tool for \gls{ML} that was developed by Microsoft. It promises end-to-end support for the complete lifecycle of the \gls{ML} application and is integrated into the cloud environment offered by Azure. Training data can be stored using the Azure Blob storage, and \gls{ML} models can be trained based on this data using clusters provided by Microsoft. Via the web interface, the developer can create a pipeline in a C\&C-like manner, but only with predefined components. AutoML techniques are also supported; for example, if data is uploaded in a CSV format, the user can specify which parts of the file's content serve as a feature and which values are to be predicted by the model. Attached to this are a fully automated construction and the training of a network.

 %1
\section{Conclusion}
\label{sec:conclusion-future-work}
In this work, we have compared two open-source \gls{mde} tools for ML-enabled software systems, namely MontiAnna \cite{KNP+19, AKK+21} and ML-Quadrat \cite{Moin+2022-SoSyM, ML-Quadrat}. First, we have conducted a case study using both tools to introduce them, as well as the concept of modeling-driven engineering of \gls{ML} software. Thereafter, we have compared the two from a functional perspective. To the best of our knowledge, this work represents the first such study focused on comparative analysis of MDE4ML in the context of IoT.

While MontiAnna has a strong focus on the development of \glspl{ann} and the integration of \gls{ML} functionality in IoT systems via MontiThings, ML-Quadrat supports the use of other \gls{ML} methods besides ANNs, and seamlessly integrates the \gls{ML} functionality and the rest of the smart IoT services. However, ML-Quadrat is limited in terms of out-of-the-box and modular support for advanced ANN architectures. Conceptually, both approaches are based on the idea of incorporating an \gls{ML} model as a component in a larger system and enabling the code to be generated, although the frameworks and languages generated may differ.

From the functional comparison conducted in this work, we can draw insights on the way model-driven engineering supports different aspects of the development of \gls{ML}-driven software. The model-driven approach relieves the \gls{ML} engineer from the burdens of solving the task in the framework-specific implementation, thus shifting the focus towards the quintessence of the development process. Thereby, it helps with the development of \gls{ML} software as well as with its integration into larger scale software systems, such as \gls{IoT} applications, through a simple specification of the key parameters without loss in flexibility. 

Both tools are research prototypes, which are still under development. However, they are provided as open-source software, and promote open standards (e.g., ONNX compatibility in the case of MontiAnna, as well as interoperability with Scikit-Learn and Keras/TensorFlow in the case of ML-Quadrat). In this way, we expect synergies and network effects in the software engineering and machine learning communities leading to a rapid adoption and extension of MDE tools for machine learning in academia, as well as their exploitation and adoption in the industry.
 %0.5

%\begin{verbatim}
  \bibliographystyle{ACM-Reference-Format}
  \bibliography{selit,refs}
%\end{verbatim}
\end{document}